\documentclass[twocolumn,showpacs,preprintnumbers,amsmath,amssymb]{revtex4}
\usepackage{graphicx}
\usepackage{dcolumn}
\usepackage{bm}

\begin{document}

\title{Fermi surface reconstruction of superoxygenated La$_2$CuO$_{4}$ with ordered oxygen interstitials}

\author{Thomas Jarlborg$^1$ and Antonio Bianconi$^{2,3,4}$}

\affiliation{
$^1$DPMC, University of Geneva, 24 Quai Ernest-Ansermet, CH-1211 Geneva 4,
Switzerland
\\
$^2$ RICMASS Rome International Center for Materials Science Superstripes, Via dei Sabelli 119A, 00185 Rome, Italy
\\
$^3$ Institute of Crystallography, Consiglio Nazionale delle Ricerche, via Salaria, 00015 Monterotondo, Italy
$^4$ Physics Department, Sapienza University of Rome, 00185 Roma, Italy}


\begin{abstract}
 
Novel imaging methods show that the mobile dopants in optimum doped La$_2$CuO$_{4+y}$ (LCO) get self-organized, instead of randomly distributed, to form an inhomogeneous network of nanoscale metallic puddles with ordered oxygen interstitials interspersed with oxygen depleted regions. These puddles are expected to be metallic, being far from half filling because of high dopant density, and to sustain superconductivity having a size in the range 5-20 nm. However the electronic structure of these heavily doped metallic puddles is not known. In fact the rigid-band model fails because of ordering of dopants and supercell calculations are required to obtain the Fermi surface reconstruction. We have performed advanced band calculations for a large supercell La$_{16}$Cu$_8$O$_{32+N}$  where $N$=1 or 2 oxygen interstitials form rows in the spacer La$_{16}$O$_{16+N}$ layers intercalated between the CuO$_2$ layers as determined by scanning nano x-ray diffraction. The additional occupied states made by interstitial oxygen orbitals sit well below the Fermi level ($E_F$) and lead to hole doping as expected. The unexpected results show that in the heavily doped puddles the altered Cu(3d)-O(2p) band hybridization at $E_F$ induces a multiband electronic structure with the formation of multiple Fermi surface spots:  a) small gaps appear in the folded Fermi surface, b) three mini-bands cross $E_F$ with reduced Fermi energies of 60, 150, and 240 meV respectively, c) the density-of-states and band mass at $E_F$ show substantial increases, and d) spin-polarized calculations show a moderate increase of antiferromagnetic spin fluctuations. All calculated features are favorable to enhance superconductivity however the comparison with experimental methods probing the average electronic structure of cuprates will require the description of the electronics of a network of multigap superconducting puddles.

\end{abstract}


\maketitle

\section{Introduction.}

The high-temperature superconducting cuprates (HTS) are heterostructures at atomic limit made of stacks of $CuO_2$ atomic planes 
intercalated by spacer layers  containing up to 25 percent of defects: oxygen interstitials, vacancies or atomic substitutions. 
The simplest cuprate La$_2$CuO$_{4}$ (LCO) is made of a single layer $CuO_2$ intercalated by La$_2$O$_{2}$ rocksalt layers. 
The Fermi surface in La$_2$CuO$_{4}$ (LCO) is formed by a single band of the $CuO_2$ layer that crosses the Fermi level \cite{pick}
. At half-filling the Fermi surface is destroyed by electronic correlations making a charge transfer gap between $Cu3d^{9}O2p^{6}$ and $Cu3d^{10}O2p^{5}$ many body electronic configurations. Insertion of dopants or defects in the spacer layers moves the chemical potential out off half-filling, forming a complex metallic phase. The rigid band model predicts a Fermi surface (FS) made of a large cylinder \cite{pick}. The breakdown of the rigid band model in the underdoped phase is shown by the appearance of i) $Cu3d^{9}O2p^{5}$ states in the charge transfer gap \cite{bianc87}; ii) a pseudo gap at the M point; iii) the Fermi surface made of short arcs \cite{dama} and iv) an electron-like small FS pocket observed by quantum oscillations and high-field Hall effect \cite{Leboeuf}.
In the underdoped regime the homogeneous metallic phase competes with electronic phase separation with the formation of a complex 
heterogeneous phase \cite{zaanen,gorkov,kugel,kresin}. The charge-spin ordering and electronic phase separation are accompanied 
by fluctuations of the local structure that diverges from the average structure 
\cite{bianc94,bianc94a,bianc94b,bianc95,bianc96,egami}.
The defects inserted in the spacer layers to dope the copper oxide electronic structure are an additional source for the 
modulation of the heterogeneous electronic states near the Fermi level \cite{little}. A particular attention has been addressed 
to oxygen interstitials that are mobile because of the misfit strain between the active and spacer layers \cite{agrestini} 
inducing a contraction of the Cu-O bonds \cite{garcia}. Controlling their ordering by thermal treatments giant effects on 
superconductivity have been reported \cite{barn,chma,geba}. Electronic structure calculations show that the displacement 
of apical-oxygen, due to oxygen defects, changes the spin-phonon coupling\cite{tj3} and the electronic states near the Fermi level in cuprates \cite{scojb}. The control of the functional electronic properties via dopants manipulation is of very high interest in manganites \cite{renn} and graphene \cite{wehl1}. 
The focus has been recently addressed to the control of structural nanoscale phase separation into dopant rich puddles embedded 
in a dopant poor background in these complex materials. The dopant rich puddles of ordered dopants has been clearly observed also 
in iron chalcogenide superconductors and it has been shown that the two structural units have a quite different electronic 
structure with a superconducting or a magnetic phase \cite{li,ricci,kseno}.
In the underdoped regime of cuprates a structural phase separation between a dopant poor antiferromagnetic phase and a dopant 
rich metallic phase with doping close to 1/8 is clearly observed in La$_2$CuO$_{4+y}$ for 0$<y<$0.055 \cite{radaelli} 
and YBa$_2$Cu$_3$O$_{6+y}$ \cite{campi}.
In the optimum and high doping regime of cuprates recent high resolution ARPES and STM experiments \cite{Vishik,lee,piriou} 
show multiple electronic components with electronic phase separation at low temperature. 
A first pseudo-gap phase with doping 
around 1/8 competes with a highly doped metallic phase with about 1/4 holes per Cu site is 
well described by band structure 
calculations, confirming previous findings of phase separation giving distinct physics beyond the superconducting dome 
\cite{bianc94a,bianc94b,bianc95}.
Superoxygenated La$_2$CuO$_{4+y}$  y$>$0.08 provides an ideal case to study the  phase separation in the optimum doping regime 
and the role of ordering of oxygen interstitials (O$_i$'s) \cite{mohn,frat,poccia,poccia2}. 
The O$_i$'s in the rocksalt LaO 
layer are mobile because of the large tensile strain in the spacer layers. This is due to the lattice 
misfit between different layers \cite{agrestini} of Cu-O bonds that is 
about 4 percent shorter than its equilibrium value of 1.97 \AA~ \cite{garcia}. Therefore, by thermal treatments it is possible 
to induce their three dimensional ordering that has been shown to enhance the critical temperature to the highest value for 
La214 \cite{poccia}. The variation of the critical temperature of the heterogeneous sample depend both on self organization 
of oxygen rich puddles and on the ordering of oxygen interstitials, as it has been shown using thermal treatments controlling 
of Tc \cite{poccia}.
Using a novel experimental microscopy, scanning nano x-ray diffraction, probing both the real and the k-space the formation 
of a granular phase in La$_2$CuO$_{4+y}$   has been discovered. Metallic heavily doped puddles, with ordered oxygen interstitials 
in the spacer layers, are embedded in a oxygen poor background \cite{frat,poccia,poccia2} at optima doping. In the heavily doped 
grains the excess oxygen interstitials are ordered as shown in Fig. 1 forming stripes along the [1,1,0] direction. 

Although such compelling pieces of evidence for the importance of oxygen order there are no theoretical calculations on the 
electronic structure reconstruction near the Fermi level in the dopant rich puddles. 
In this work we present electronic structure results for the highly hole doped puddles of ordered oxygen interstitial 
in superoxygenated La$_2$CuO$_{4}$.
The method of calculation based on experimental structural information consider the supercells of O-rich LCO in sect. II. 
In sect. III we discuss the results of the calculations, and some ideas for future works are given together with the 
conclusions in sect. IV. 

\begin{figure}
\includegraphics[height=12.0cm,width=8.0cm]{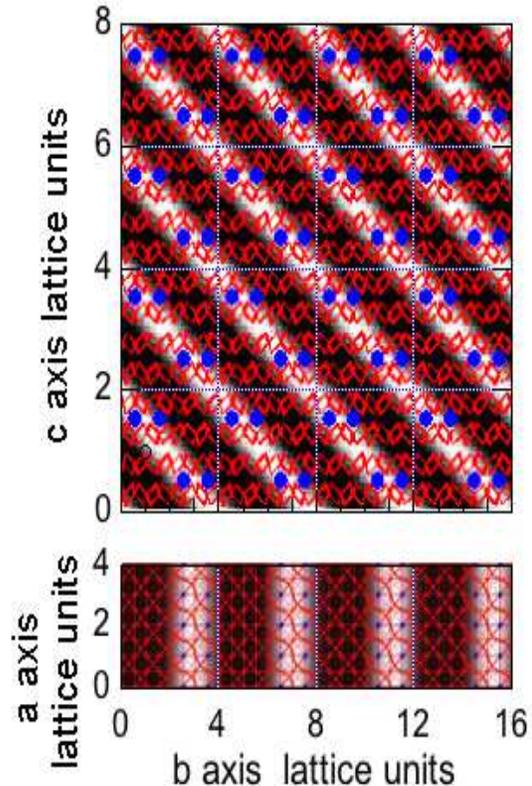}
\caption{(Color online) The ordering of oxygen interstitials (blue solid dots) in La214 in the (b,c) plane 
(upper panel) and in the (a,b) plane of the Fmmm structure as determined by x-ray diffraction 
using synchrotron radiation.  The unit cell is La$_{16}$Cu$_8$O$_{32+N}$ with $N$=2.}
\label{fig1}
\end{figure}

\section{Method of calculation.}

The calculations are made using the linear muffin-tin orbital (LMTO) method and the
local spin-density approximation (LSDA).  Electronic correlation can be disregard because the oxygen rich puddles very far from half filling. 
The details of the methods have been published earlier 
\cite{tj1}-\cite{tj11}. The elementary cell of La$_2$CuO$_4$ (LCO) contains La sites 
at (0,0,$\pm$.721c), Cu at (0,0,0), planar O's at (0.5,0,0) and (0,.5,0) and apical
O's at (0,0,$\pm$.358c), in units of the lattice constant $a_0$=3.8 \AA, where c=1.24.
In addition to the MT-spheres at the atomic sites
we insert MT-spheres at positions (.5,0,$\pm$.5c) and (0,.5,$\pm$.5c) to account for the positions
of empty spheres. 
The atomic sphere radii are 1.76 \AA~(La), 1.26 \AA~ (Cu), 1.18 \AA~ 
(planar O, O$_i$ and empty spheres), and 1.21 \AA~ (apical O), respectively.
The z-projection of the corresponding irreducible Brillouin zone (BZ) is shown in Fig.
\ref{fig4}a by the limiting points $\Gamma-X-M$, whereas the IBZ for the doubled cell (La$_4$Cu$_2$O$_8$)
with ordered antiferromagnetic (AFM) order
would be limited by $\Gamma-X-M"$. 
Eight units of the elementary cell La$_2$CuO$_4$ are put together to form a long supercell, La$_{16}$Cu$_8$O$_{32}$,
with the generating
lattice vectors $(1,-1,0), (4,4,0)$, and $(.5,.5,c)$, so that 
its axis is oriented parallel 
to the (1,1,0)-direction, i.e. at 45 degrees
from the Cu-O bond direction along (1,0,0). 
The band calculations are made for this supercell containing 72 sites totally, where one or two of the empty
sites at the interstitial positions are occupied by excess oxygen ions. 
These supercells are chosen in order to represent fairly well the experimentally determined structures \cite{frat}.
A small lattice relaxation
near oxygen interstitials is included in the first elementary segment of the
supercell, so that apical-O are pushed away about $0.03 a_0$ in the x-y plane.
The basis set goes
up through $\ell$=2 for all sites. The z-projected IBZ corresponding to the supercell is shown
in \ref{fig4}a by the limits $\Gamma-M_3-X_3-R$. Self-consistency is made with 192 k-points and final
results are based on 408 points in the IBZ. 

No interstitial sites are occupied with O in
one set of the calculations (called LCO-0). 
The DOS of the LCO-0 calculations agree well with other published results on LCO \cite{pick}. 
The LCO-0 results serve as a reference for comparison with the results with oxygen interstitials.
For instance, the FS for (lightly doped)
LCO forms a circle centered at the $M$ point in BZ of the elementary cell, but since the
BZ corresponding to the supercell is folded and very flat along (1,1,0) it is necessary
to identify the circle from several FS pieces in the folded zone. 
In two other sets of calculations we insert one (La$_{16}$Cu$_8$O$_{32+1}$, LCO-1) 
or two (La$_{16}$Cu$_8$O$_{32+2}$, LCO-2) oxygen ions at
interstitial positions. In LCO-2 the two O are put at equivalent positions in two
adjacent AFM cells,
so that two stripes of O$_i$'s, separated by $1.41 a_0$
are formed along $\vec{x}$ (and $\vec{z}$) \cite{frat}.

The excess O$_i$'s sit at the interstitial interlayer positions,
above the oxygen ion in the CuO$_2$ plane of the  
the orthorhombic unit cell, and form 3D ordered puddles below 350K, c.f. Fig. \ref{fig1}.
The insertion of O$_i$'s is expected to induce hole doping, because  
each new oxygen interstitial will bring 4 new bands well below $E_F$ (one "s" and 3 "p"), but the oxygen has 
only 2 "s" and 4 "p" electrons. Therefore, simple arguments suggest that one Cu-O  band becomes unfilled, i.e. 
$E_F$ has to go down relative to the rest of the bands. However, this picture gets complicated by the fact that 
other atoms like La serve as charge reservoirs, lattice reconstructions are likely, and in addition the excess O 
positions are ordered in stripe-like patterns \cite{frat}.

It is often argued that correlation is too strong for having traditional bands in undoped cuprates
when the Cu-d band is half-filled \cite{???}. On the other hand, ARPES (angular-resolved photoemission
spectroscopy) and ACAR  detect FS's and
bands that evolves with doping and agree well with DFT (density-functional theory)
calculations \cite{pick,dama,posi}. Therefore,
since we study doped cases here, it is not likely that correlation is more important than what is already
included in our DFT calculations.

\begin{table}[ht]
\caption{\label{table1}
Total and band decomposition of the DOS at $E_F$ in units of $(cell \cdot eV)^{-1}$, 
and average charges
within each MT-sphere of the different atoms, in units of valence electrons per
atom. Only 3 bands have a significant DOS at $E_F$, see the text for the numbering of the
3 bands in each case. The charges on Cu and especially La differ from site to site
depending on the proximity to interstitial O. The O-charges are averaged planar and apical O
and do not include the O$_i$'s. 
  }
  \vskip 2mm
  \begin{center}
  \begin{tabular}{l c c c c c c c}
  \hline
    & N$(E_F)$ & low & mid & upper & $Q_{Cu}$ & $Q_{La}$ & $Q_{O}$ \\
  \hline \hline

 LCO-0  & 8 & 1.5 & 3.0 & 4.2 & 10.23 & 7.86 & 6.43 \\
 LCO-1  & 20 & 3.5 & 4.8 & 12 & 10.19$\pm$.02 & 7.92$\pm$.14 & 6.43 \\
 LCO-2  & 25 & 3.0 & 3.7 & 18 & 10.15$\pm$.02 & 7.98$\pm$.13 & 6.42 \\

  \hline
  \end{tabular}
  \end{center}
  \end{table}

\begin{figure}
\includegraphics[height=8.0cm,width=9.2cm]{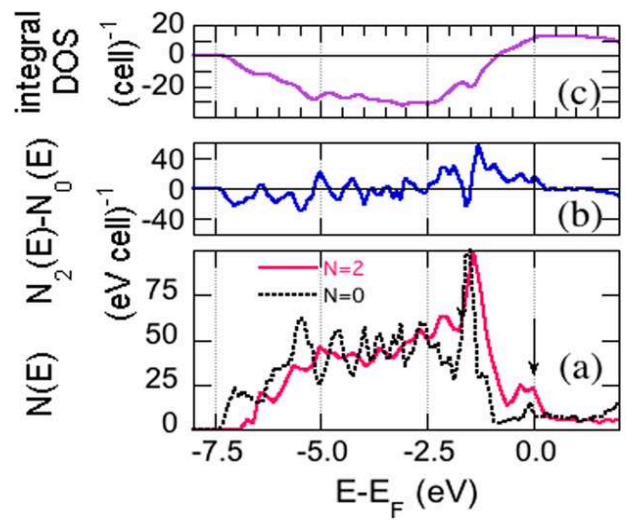}
\caption{(Color online) Lower panel: The total DOS for La$_{16}$Cu$_8$O$_{32+N}$ with $N$=0 and $N$=2. 
A broadening with a FWHM of $\sim$ 0.01 eV is included.
Middle panel: The difference of the DOS for $N$=2 minus the DOS for $N$=0. 
The maximum at -2 eV indicates the main energy location of the O(2p) impurity states of oxygen interstitial. 
Upper panel: The number of states, i.e. the integrated value of what is shown in the middle panel.}
\label{fig2}
\end{figure}

\section{Results and discussion.}

\begin{figure}
\includegraphics[height=8.5cm,width=10.0cm]{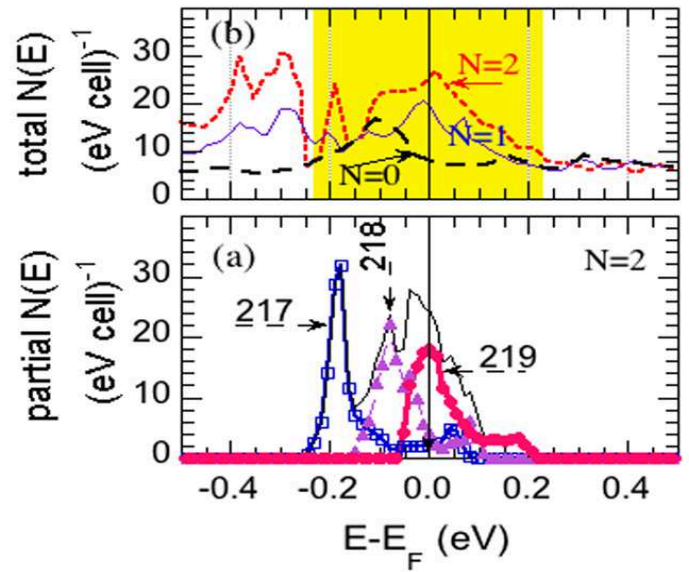}
\caption{(Color online) Upper panel: The total DOS for different bands near $E_F$ in La$_{16}$Cu$_8$O$_{32+N}$ 
for  $N$=0,1 and 2, with a broadening of about 5 meV. Lower panel: The partial DOS of the 3 minibands 
crossing $E_F$ in La$_{16}$Cu$_8$O$_{32+2}$ and their sum (thin line) with broadening of about 3 meV. 
The Fermi energy of the upper miniband (219) is only 60 meV, to be compared with a superconducting gap of the order 
of 30 meV. The Fermi energy of the other two minibands (218,217) are 150 meV and 240 meV. 
The Fermi energy of the corresponding bands (213-211 in the undoped case) for $N$=0 is about
180, 300 and 950 meV, respectively. 
The gap between the 219 and 218 minibands, about 80 meV, is of the same order of magnitude as
the so called pseudo gap 
in the anti-nodal point of the Fermi surface.
}
\label{fig3}
\end{figure}

 The nonmagnetic (NM) total DOS for La$_{16}$Cu$_8$O$_{32+N}$ near the Fermi level are shown 
 in Figs. \ref{fig2}-\ref{fig3}. 
 It can be seen that the DOS is increasing considerably near
 $E_F$ when one or two O$_i$'s are added in form of stripes. The DOS difference between the cases $N$=0 and $N$=2
 in Fig. \ref{fig2}b shows that most of the  O$_i$-p bands fall in the interval [-2.5,0] eV. Strong
 hybridization puts new states also on atoms around the O$_i$ sites. The integrated differences (as the one shown in 
 Fig. \ref{fig2}c) show that the electron count is about the same for all three cases at approximately -1.0 eV below
 the respective $E_F$'s.
The DOS at $E_F$, N$(E_F)$, increases roughly three times when two O are added, see Table \ref{table1}. 
In the undoped sample the Fermi level $E_F$ is well above the small peak in the DOS associated 
with the van-Hove singularity.
In the doped cases the peak in the total DOS splits forming two or three peaks 
separated by a partial gap slightly below $E_F$, so that $E_F$ is tuned exactly at the top of one the DOS peaks.
The DOS functions from individual bands near $E_F$ show similar shapes, but within a narrower energy range when
the number of O$_i$'s increases. The partial band DOS from bands 217-219 for $N$=2 are displayed in Fig. \ref{fig3}b;
The shapes of the DOS from bands 211-213 for $N$=0 look similar, but are $\sim$3 times wider.
Only three bands have a significant
N$(E_F)$, see also Table \ref{table1}.
Of the two bands that contribute most to the total $E_F$, are the mid- and upper-band, 218 and 219 for $N$=2
(and 212-213 for $N$=0, and 215-216 for $N$=1). 
These upper bands hybridize with O$_i$ and make the local DOS on the
added O very high, about twice as high as the local d-DOS on Cu. The local
Cu-d DOS is largest on Cu-sites near the O$_i$'s
(typically twice the DOS on Cu far from O$_i$). On the average there is a quite delocalized
increase of the Cu-d DOS coming from the uppermost band when oxygen interstitials  are added.
This increase is smaller than of the total DOS, because of the large O-p DOS, but it is most
striking for the upper band, suggesting that this band is most important for 
any benefit of superconducting properties.  

The changes of the effective charges (see Table \ref{table1}) show a weak hole doping on Cu when the
number of O$_i$ increases. A weak doping is also seen on the planar O charges in agreement with experiments showing only about one hole per Cu site induced by each oxygen interstitial at high doping. The charge on O$_i$ is $\sim$0.15 electrons less than within an apical O of the same volume. This difference makes the 1s core levels to go
up by about 1.5 eV on O$_i$ compared to the other O.

\begin{figure}
\includegraphics[height=15.0cm,width=8.5cm]{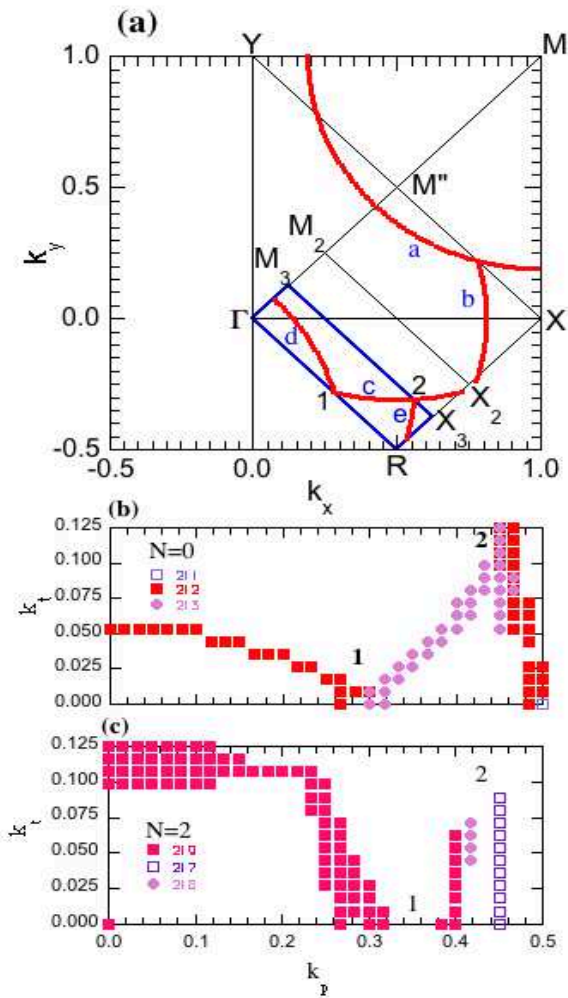}
\caption{(Color online) Upper panel (a): A schematic view of how a circular M-centered FS in the normal BZ
(the irreducible part is limited by $\Gamma$-X-M)
is folded into the BZ of the supercell given by the rectangle $\Gamma$-M$_3$-X$_3$-R.
The  M" is halfway between $\Gamma$ and M and $\Gamma$-X-M" is the limit of the basic AFM BZ.
Middle panel (b); the FS pieces (bands 211-213) in the $k_z$=0 planes of the irreducible BZ
(i.e. within $\Gamma$-M$_3$-X$_3$-R)
for $N$=0.
Lower panel (c); the FS pieces (bands 217-219) in the $k_z$=0 planes of the irreducible BZ
for $N$=2.
}
\label{fig4}
\end{figure}

The FS of undoped LCO consists of a rather 2-dimensional almost circular cylinder
with origin at the M-point. The radius of the cylinder is quite large so that the FS
englobes the M"-point in the center, see the schematic FS in Fig. \ref{fig4}a. The radius
gets larger for increasing doping and the FS reaches the X- and Y-points at optimal
doping when $E_F$ coincides with the van-Hove singularity peak in the DOS. AFM order
on Cu sites leads to a cell doubling and the BZ is folded along the X-Y line
to obtain the irreducible part of the AFM-BZ within $\Gamma$-X-M". One portion of 
the FS will be folded into the section "b" in Fig. \ref{fig4}a, while "a" is already
within the AFM-BZ. Our band calculations are made for a 4 times extension of the
AFM cell in the (1,1,0) direction, and the corresponding BZ is found from two more
foldings of the AFM-BZ, the first at the line M$_2$-X$_2$ and once more at M$_3$-X$_3$.
The new irreducible BZ is confined within $\Gamma$-M$_3$-X$_3$-R, and the portions "d",
"c" and "e" constitute the down-folded FS circle of the elementary BZ into the BZ for the supercell. 
These 3 bands are found
at $E_F$ in the supercell calculations for LCO-0 and LCO-2 as can be seen in Figs. \ref{fig4}b-c. 
Fluctuations tend to open a
gap at the point "1" \cite{tj11,harr,meng}. 
It is also realized that portions "e" and parts of "c" 
will be gone at high doping. This is because the original $M$-centered FS circle will increase its
radius for higher doping; the FS crossing that occurs on the $M-X$ line for low doping will
reach the $X$-point at optimal doping, and ultimately cross the $\Gamma-X$ line when the doping is
increased further. The traces of the "e" and "c" portions will also be gone if there is a gap of the original FS at X,
i.e. if the FS circle in the elementary BZ breaks up into a banana-shaped feature centered at $M"$.

In Figs. \ref{fig4}b-c are displayed the FS pieces in the $k_z$=0 planes of the irreducible BZ
for $N$=0 and 2.
For the calculations without interstitial O it is easy to recognize the FS-model in the 
real FS in the $k_z$=0 plane. There are minor modifications in other $k_z$ planes because
of a weak 3-D dispersion.
The energy window is set narrower for the results with 2 interstitial O in order to
avoid too many points for the more flat bands in this case. Many of the same structures of
the FS are seen in the plots for the results with interstitial O as for the results without them.
But two differences draw our attention: First, there are many FS-points (this is more striking
when the same energy window is used as for the undoped cell) to the left of the panel, close to $\Gamma$, when
there are O-interstitials,
which suggests that mainly the FS piece "d" in Fig. \ref{fig4} has gained a higher DOS. This agrees with the high DOS
for the upper band (219 for $N$=2). In addition, section "d" 
is closer to the M$_3$-X$_3$-line,
indicating hole doping as for an enlarged unfolded FS circle. Secondly,
no FS is seen to the right (i.e at $k_p$=0.5) for $N$=2, which probably means that
a gap has opened near the X-point, or that the effective doping is such that the
original unfolded FS already is beyond the X-point. A separate small hole pocket appears  
near $k_x \sim$ 0.45. An effort to force a back folding of that pocket into the original zone 
would put that band close to X on the
M"-X line. 
The possibility of a gap at $X$ is interesting, since a pseudogap is often associated
with enhanced $T_C$ near optimal doping
for conventionally doped LSCO, and a small pocket nearby sets up the condition for the Lifshitz
mechanism \cite{inno1}. 
Such band features can to some extent be
understood; A dispersive band in the [1,1,0]-direction breaks up by the new
periodicity and the band becomes flat at the limit of the (shortened) BZ where even small gaps 
might appear. The DOS will be large if the band remains flat in other k-directions. It is a matter of
tuning the periodicity to the doping so that $E_F$ coincides with a favorable structure in the DOS \cite{tjapl}.
It has been demonstrated that the pseudogap depends on stripe-like
potential modulations in the (1,0,0)-direction, as from phonon or spin waves \cite{tj1,tj7,tj6},
but here the stripes are oriented in the (1,1,0)-direction. These features are a bit different
when only one oxygen interstitial is added.

A high DOS is normally connected with strong spin fluctuations. It has been shown that
for a certain correlation between the wave length of spin fluctuations and the doping, the pseudogap
can open right at $E_F$, where also phonons can contribute \cite{tj3,tj6}.  
Now the modulation is in the (1,1,0)-direction without 
optimization of the length of the super cell to the doping, but it is still of interest
to see if AFM spin fluctuations can become stronger by interstitial O. This is seen
from spin-polarized calculations where AFM fields are applied to Cu atoms to create a
spin wave along the cell. Two pairs of Cu near and far from the O$_i$'s have
the applied fields ($\pm 5 mRy$), while other Cu at the nodes of the wave have no
applied field. The local exchange enhancement (Stoner factor) and moments will be the measure
of the strength toward spin fluctuations.

\begin{table}[ht]
\caption{\label{tabspin}
Local Stoner enhancement, S, and within parentheses the induced local
moment (m in units of $\mu_B/atom$) on 4 Cu calculated for AFM fields of $\pm 5 mRy$.
The two first Cu are closest to the interstitial O, while Cu(3) and Cu(4)
far from them are most disposed to show magnetism.  Distortion in one layer of 
apical O and unequal distances to interstitial O in LCO-2 contribute to differences between Cu sites. 
  }
  \vskip 2mm
  \begin{center}
  \begin{tabular}{l c c c c }
  \hline
    & Cu(1) & Cu(2) & Cu(3) & Cu(4)  \\
  \hline \hline
        & S~~~  (m)  & S~~~  (m)    & S~~~ (m)   &  S~~~ (m)    \\
  \hline
 LCO-0  & 1.46 (.061) & 1.46 (-.060) & 1.46 (-.060) & 1.46 (.060)  \\
 LCO-1  & 1.51 (.063) & 1.42 (-.050) & 1.62 (-.079) & 1.63 (.081) \\
 LCO-2  & 1.45 (.060) & 1.45 (-.059) & 1.43 (-.055) & 1.56 (.071) \\

  \hline
  \end{tabular}
  \end{center}
  \end{table}
The results are summarized in Table \ref{tabspin}. As seen, with extra O$_i$'s the tendency is such that the local
enhancements are larger on sites far from the oxygen interstitial, on sites where the DOS is not
highest.
The DOS on Cu near oxygen interstitial is increased by direct
hybridization with p-states on those O-sites. 
The local O-p DOS near $E_F$ is large on O$_i$ (larger than 1 $eV^{-1}$)
and nearby apical O (0.5-1 $eV^{-1}$), which is in contrast to the case without interstitial O or far away
from the O$-i$ where the DOS is small (less than 0.1 $eV^{-1}$). This fact resists to an enhancement on Cu, 
because the Cu-d and O-p
bands are tightly bound and the latter are less adapted for magnetism.

\section{Conclusion.}

Oxygen interstitials added in form of stripes in the spacer layers form three overlapping mini-bands 
crossing $E_F$ making the DOS larger 
and $E_F$ appears at the maximum of one of the DOS
peaks. Thus, the natural growth of O$_i$-stripes in the oxygen rich puddles seems to be an efficient way to
generate potential modulations for having $E_F$ on a DOS peak with high
N$(E_F)$, accompanied by enlarged electron-phonon coupling and $T_c$ \cite{tjapl}.
There are many possibilities for spectroscopic verification of the local electronic
structure near the interstitials.
It would be very important to perform ARPES measurements using photon excitation beams focused on nanoscale spots to 
measure the electronic structure of the super-oxygenated puddles calculated here. It is clear that ARPES experiments 
using large photon beams give information on the spatially averaged structure probing at the same time both oxygen-rich 
and oxygen-poor domains. Therefore we propose that nano-focused photoemission experiments would be 
a powerful probe to detect band narrowing in oxygen ordered
O$_i$'s. The fact that we find an upward shift of the O-1s core level of the order 1.5 eV
on the single stripe of O$_i$ and nearby apical O (compared to the O far from the interstitials) suggests that the increase
of local O$_i$-p DOS can be detected via the threshold energies of soft X-ray emission and/or absorption
spectroscopy. The more than tenfold increase of the local p-DOS near $E_F$ on interstitials 
is a large effect that should modify spectroscopic intensities as function of increased oxygen content.

The changes of electronic structure on Cu sites are strong and moderately favorable to AFM spin fluctuations with
possibilities for enhanced $\lambda_{spin}$. Furthermore, there is a segmentation of the 
original FS with a superconducting puddles, into smaller disconnected pockets, from which the shape resonance in the 
superconducting gaps near a Lifshitz transition emerges as an additional mechanism for high T$_c$ \cite{inno1,inno2}.
We show that the ordered lattice with 3D oxygen interstitials in oxygen rich puddles in LCO provides a realization 
of a multi-band metallic puddles.
All calculated features of the electronic structure of oxygen rich nano-puddle are expected to be favorable 
to enhance superconductivity however the comparison with experimental methods probing the average electronic structure 
of cuprates will require the study of networks of superconducting multi-band puddles described in this paper. 
Finally theoretical works on HTS focusing on granular superconductors \cite{demello2,ginestra12}, where  
the maximum superconducting critical temperature occurs at a critical density for the superconductor-insulator 
transition in a network of 2d percolation clusters  \cite{ginestra13} could investigate networks of multi-band 
superconducting puddles for understanding many puzzles of high temperature superconductors.

\end{document}